# Magnetic Structure of Hexagonal Mn-doped LuFeO$_3$

Steven M. Disseler[1], Xuan Luo[2], Yoon Seok Oh[3], Rongwei Hu[3], Dylan Quintana[4], Alexander Zhang[1,*], Jeffrey W. Lynn[1], Sang-Wook Cheong[2,3], William Ratcliff II[1]

**Affiliations**

1. NIST Center for Neutron Research, National Institute of Standards and Technology, Gaithersburg, Maryland 20899, USA
2. Laboratory for Pohang Emergent Materials, Pohang University of Science and Technology, Pohang 790-784, Korea
3. Rutgers Center for Emergent Materials and Department of Physics and Astronomy, Rutgers University, Piscataway, New Jersey 08854, USA
4. Carnegie Mellon University, Pittsburgh Pennsylvania 15213, USA
*Montgomery Blair High School, Silver Spring Maryland 20901, USA

**Abstract**

Neutron scattering techniques are used to investigate the crystalline and magnetic structure of LuFe$_{0.75}$Mn$_{0.25}$O$_3$ in bulk polycrystalline form. We find that the crystalline structure is described by the hexagonal *P6$_3$cm* space group similar to that of thin-film LuFeO$_3$, and that the system orders antiferromagnetically below $T_N$ = 134 K. Inelastic neutron scattering reveals nearest-neighbor superexchange parameters that are enhanced relative to LuMnO$_3$. The observation of significant diffuse scattering above $T_N$ demonstrates the frustrated nature of the system; comparisons with similar materials suggest the ground state magnetic configuration is sensitive to local crystallographic distortions.

**Introduction**

Multiferroic materials, which have both ferroelectric (FE) and magnetic order, have been of increasing interest in recent years for novel electronic and spin-based devices and as a tool to study the more fundamental aspects of the coupling between spins and the lattice [1]. Although several material families have been discovered which demonstrate multiferroic behavior in this way, relatively few systems exhibit both FE and magnetic order at or above room temperature, with BiFeO$_3$ the best-known example [2]. The hexagonal manganites *R*MnO$_3$ (*R = La-Lu, Y, etc.*) are one class of intensely studied multiferroic systems, as FE order develops well above room temperature—above $T_c \sim$ 1000 K in some cases—resulting from a crystalline distortion which induces the necessary $P6_3cm$ polar structure [3, 4] shown in Fig. 1(a). Magnetism in these systems is only weakly coupled to the FE transition, as Mn spins order antiferromagnetically with $T_N \leq 100$ K, forming a frustrated noncollinear 120° structure within each *a-b* plane, an example of which is shown in Fig. 1. On the other hand, it has recently been suggested that a very similar class of materials, the hexagonal ferrites, such as $LuFeO_3$ (h-LFO), may exhibit greatly improved magnetic properties relative to their manganite counterparts [5, 6] due to the enhanced exchange interaction and larger moments of the $Fe^{3+}$ relative to $Mn^{3+}$. This includes higher magnetic ordering temperatures of the transition metal (TM) moments and larger magnetoelectric effects in the bulk, a necessary property for actual device realizations [6, 7].

Recent investigations of thin films of h-LFO suggest some evidence for magnetic order at much higher temperatures [8], possibly even above room temperature [6], as well as the appearance of a net ferromagnetic (FM) moment. However, determination of the complete magnetic structure as well as studies of low energy excitations through inelastic neutron scattering are generally unfeasible for such small material volumes. Similar measurements of h-LFO in bulk form are therefore very important to determine unambiguously the ground state magnetic configuration and to discern any extrinsic or purely morphological effects that may arise in films.

Bulk $LuFeO_3$ orders antiferromagnetically with the *C*-type structure around 600 K, well above room temperature [9]. However, the crystal structure that forms is nominally the non-polar orthorhombic *Pbnm* structure, precluding ferroelectricity. Fortunately, the required polar $P6_3cm$ structure is a metastable phase that can be stabilized through either morphological constraints or through off stoichiometric growth [10, 11]. Another approach, as demonstrated in this work, is to introduce Mn on the Fe site in LFO, as $LuMnO_3$ is stable in the hexagonal form in the bulk [3]. Here we focus on one such compound, $LuFe_{0.75}Mn_{0.25}O_3$ which is found in the hexagonal $P6_3cm$ structure necessary for multiferroic behavior. We find that long-range antiferromagnetic order of the TM species occurs below $T_N$ = 134(1) K that can be described by the $\Gamma_2$ representation alone. Inelastic neutron scattering measurements reveal the existence of conventional spin wave excitations in the ordered state which are well described by an antiferromagnetic Heisenberg model with an in-plane nearest neighbor exchange $J_{nn}$ =-5.3 ± 0.01 meV, and a very small $J_c$ = -0.05 ±0.02 meV along the *c*-axis between planes. A small single ion anisotropy energy of $D_z$ = 0.06 ±0.01 meV is also found. Moreover, the distribution of diffuse scattering suggests that different magnetic representations have very similar energies such that variations in the observed ground state between bulk and thin films may be the result of subtle differences in the local environment of the transition metal species.

**Experiment**

A five gram polycrystalline sample of $LuFe_{0.72}Mn_{0.28}O_3$ (h-LFMO) was obtained by the solid-state reaction method. The resulting powders were characterized by x-ray diffraction and were found to be single phase and homogeneous within the resolution limits of the instruments used. Magnetic susceptibility was measured between 2 K and 400 K under both zero field cooled (ZFC) or field cooled (FC) conditions in a field of 0.2 T. Neutron scattering measurements were performed at the NIST Center for Neutron Research on the BT-1 powder diffractometer and on the BT-7 [12] and SPINS triple-axis spectrometers. The sample was sealed in an Al canister with helium exchange gas and was cooled to 5 K in a closed-cycle refrigerator or to 2 K in a helium cryostat. The high resolution neutron powder diffraction (NPD) patterns were taken on BT-1 at 200 K and 5 K using neutrons with wavelength $\lambda$ = 1.540 Å and $\lambda$ = 2.0775 Å produced by Cu(311) and Ge(311) monochromators, respectively. Rietveld refinement of the crystalline and magnetic structure was performed with FULLPROF [13].

Measurements of the diffuse scattering were performed on BT-7 using the position sensitive detector (PSD) in two-axis mode with initial energy $E_i$ = 13.7 meV, pyrolytic graphite (PG) monochromator and horizontal collimations full-width-at-half-maximum (FWHM) $open - M - 80' - S - 80' - radial - PSD$. Inelastic measurements were taken using final energy $E_f$ = 14.7 meV with PG monochromator and analyzer with $open-80'-80'-120'$ collimations, while on SPINS horizontal collimations of $guide-80'-80'-open$ with $E_f$ = 5 meV and cooled Be-filter. Uncertainties where indicated are statistical in origin and represent one standard deviation.

**Results and Discussion**

The crystal structure of h-LFMO was refined from the NPD pattern shown in Fig. 2(a) in the paramagnetic region at $T$ = 200 K. All peaks could be indexed by the hexagonal *P6$_3$cm* space group (no. 185) using lattice parameters $a = b$ = 5.993(1) Å, $c$ = 11.634(1) Å, with the exception of those from the Al holder which were excluded. We find a goodness of fit, $\chi^2$ = 1.8 and a TM displacement along the *a*-axis of $x/a$ = 0.326± 0.002. The Mn occupancy was found be to be 0.279(3), slightly enhanced relative to the as-prepared value 0.25. The crystal structure is shown in Fig. 1(a) where the TM ions are shown in the center of the tilted oxygen bi-pyramids. The FE polarization calculated from the refined displacements of the Lu and O ions is found to be $P$ = 5.1(1) μC/cm$^2$ comparable to that measured in other *P6$_3$cm* materials such as YMnO$_3$ ($P$ = 5.5 μC/cm$^2$) [14].

A sharp transition in the temperature dependent magnetic susceptibility(Fig. 3) is observed at $T_N$ = 135 K, below which the susceptibility is strongly hysteretic with a small net ferromagnetic moment evident in the field-cooled case. Interestingly, this net moment does not saturate, and a maximum in the field-cooled susceptibility is observed near 90 K but remains non-zero even at 2 K. The inverse susceptibility (not shown) follows a Curie-Weiss law above 300 K from which a negative Weiss temperature $\theta_W \approx$ -1100 K is extracted. The ratio of these temperatures scales, $|T_N/\theta_w| \approx 0.1$, is similar to that found in h-*R*MnO$_3$ systems and is indicative of the two-dimensional nature of the system and inherent geometrical frustration of the triangular lattice.

The ground state magnetic structure was determined from NPD at 5 K (Fig. 2(b). Here, new Bragg peaks (*h*, 0, *l*), *l* odd, are observed at positions forbidden by the *P6$_3$cm* space group. These peaks may be indexed using a magnetic cell identical to the chemical which defines a propagation vector $k = 0$. Potential magnetic structures were then generated following the representational analysis described in previous work [15-18]. There are four one-dimensional irreducible representations, denoted $\Gamma_1$ to $\Gamma_4$ as shown in Fig. 1 (b)-(e). The $\Gamma_1$(b) and $\Gamma_3$(d) representations form a homometry (Bragg intensities are identical for each representation and therefore cannot be distinguished), as do $\Gamma_2$(c) and $\Gamma_4$(e). The homometric sets can be differentiated by the parity dependence of the magnetic interaction vector for reflections in the (h, 0, l) scattering plane [16]. This results in significant intensity at the (1, 0, 0) reflection in the ordered state for $\Gamma_1$ and $\Gamma_3$ representations, with no such intensity for $\Gamma_2$ or $\Gamma_4$. In Fig. 2(b). We find no change in intensity at the (1, 0, 0) reflection between 200 K and 5 K, indicating that the ground state must be given by either the $\Gamma_2$ or $\Gamma_4$ representations, in contrast to an admixture of representations found in thin films [5]. There are two two-dimensional representations allowed by the space group symmetry which are not considered here as they give rise to ferromagnetic structures in the a-b plane [18], inconsistent with the weak ferromagnetism observed in magnetic susceptibility measurements. The lineshapes of the magnetic peaks are well described by the resolution-limiting standard pseudo-Voigt function indicating that the magnetic order is long range both within the hexagonal plane and along the c-axis. In particular no evidence of a Warren lineshape, which would indicate a short correlation length along the c-axis, was observed [19].

The magnetic structure was refined using the $\Gamma_2$ representation. The basis vector corresponding to an out-of-plane ferromagnetic moment was fixed to be zero as the FM moment as determined from susceptibility is far too small to be detected in a powder diffraction experiment. From this refinement, an ordered magnetic moment $m = 3.75(1)$ $\mu_B$/ion was found using a single magnetic moment to represent both Fe and Mn moments, resulting in a goodness of fit $\chi^2 = 2.4$. This moment is significantly reduced relative to that expected for a simple mix of $Fe^{3+}$ where ($g\sqrt{S(S+1)} = 5.9$ $\mu_B$/Fe) and $Mn^{3+}$ (4.9 $\mu_B$/Mn), but is consistent with that observed in numerous studies of similar hexagonal $R$MnO$_3$ materials [16, 18, 20].

The integrated intensity of the magnetic (1, 0, 1) reflection was measured as a function of temperature on SPINS as shown in the inset of Fig. 3. The solid curve is a simple mean field fit with $T_N$ = 134(1) K, in good agreement with that observed in magnetic susceptibility and almost twice that of LuMnO$_3$ [20]. The observed onset of magnetic order in bulk is much lower than that claimed in very thin films [5], but it is in good agreement with recent measurements of thicker films [8].

The energy-integrated scattering over a range of $Q$ including the (1, 0, 0) and (1, 0, 1) reflections as shown in Fig. 4(b). We find a region of diffuse scattering below 150 K centered near $Q$ = 1.2 Å$^{-1}$ corresponding to the (1, 0, 0) reflection, which vanishes for temperatures $T < T_N$. Correlated paramagnetic scattering above $T_N$ has been observed at this reflection in YMnO$_3$ [21], and is due to inherent frustration of the lattice resulting in fluctuations of small AFM clusters, as commonly found in geometrically frustrated systems [22]. However, YMnO$_3$ eventually orders with the $\Gamma_1$ representation [18] in which the (1, 0, 0) reflection is allowed, unlike what we have found here for h-LFMO where scattering about the forbidden reflection is unexpected. The presence of enhanced scattering at both (1, 0, 0) and (1, 0, 1) reflections above $T_N$ suggests that the system exhibits correlations related to both $\Gamma_1$ and $\Gamma_2$ representations. This is particularly interesting given the observation of a combined $\Gamma_1 + \Gamma_2$ ground state in thin-films of h-LFO, which corresponds to an overall rotation of the moments by an angle $\phi$ [15]. A similar phenomena is observed in ScMnO$_3$, which orders with the $\Gamma_2$ representation near 130 K before undergoing a spin reorientation to a combined $\Gamma_1 + \Gamma_2$ below $T$ = 80 K [18]. Taken together these results suggest that these two representations are quite close in energy, and that subtle variations in the local crystal structure may drive the system from one to the other or stabilize an intermediate form. Indeed, recent first principles calculations of h-LFO indicate that the energy of the $\Gamma_1$ structure lies above the ground state $\Gamma_2$ structure by an amount smaller than the single ion anisotropy [6].

We now turn attention to measurements of the spin wave excitations in the ordered state The presence of spin wave excitations is inferred from the large density-of-states at small wave vectors with rapidly decreasing intensity at higher $Q$, as seen in $I(Q)$ for constant energy transfer scans at $E$ = 3, 6, 9, and 12 meV shown in Fig. 5(a)-(d) respectively. Constant $Q$ = 1.3 Å$^{-1}$ scans, corresponding to the wave vector of the (1,0,1) magnetic Bragg peak, are shown in Fig. 5(e) and reveal that much of the spectral weight is pushed above $E$ ~ 5 meV in the ordered state at low temperatures, indicative of a spin-gap in the ordered state originating from single-ion anisotropy. Detailed measurements of the low-energy scattering for several temperatures in Fig. 5(f) show that this gap gradually closes as the temperature approaches $T_N$ and that the spectral weight shifts to quasi-elastic scattering representing simple spin diffusion in the paramagnetic state.

We use a simple Hamiltonian, $H = -\sum_{\langle ij \rangle} J_{ij} S_i S_j + \sum_i D_z \left( S_i^z \right)^2$ for the exchange interactions along several important pathways to model the observed spin wave density of states. We include the superexchange interaction between nearest neighbor moments within each plane $J_{nn}$, a much weaker inter-plane exchange $J_c$, and a single-ion anisotropy term $D_z$. We constrained $J_c < 0$ to fix the magnetic ground state in the $\Gamma_2$ representation, but it should be noted that because of the smallness of this exchange value the overall dispersion does not change significantly if $J_c > 0$ and the $\Gamma_4$ state is used instead [23]. A more complex Hamiltonian including further neighbor interactions and the Dzyaloshinskii-Moriya interaction could not be reliably studied given the polycrystalline nature of our sample and consequent broadness of the main features in the data.

The dispersion was then calculated and powder averaged using the SpinW package [24] and convoluted with the instrument resolution function. The resulting spectra could be fit to the data in Fig. 5(a-e) to extract the exchange and anisotropy parameters. A constant background term and uniform scaling factor were also included as fitting parameters. The resulting best fits are shown by the solid (red) curves where $J_{nn}$ =-5.3 ± 0.01, $J_c$ = -0.05 ± 0.01 meV, and $D_z$ = 0.06 ± 0.01 meV. This parameter set captures many of the salient observed features, including the appropriate locations and intensity of the scattering distribtution as well as the size of the spin-gap in the ordered state.

This $J_{nn}$ is over 25 % larger than that determined from single crystal measurements of LuMnO$_3$ [23], and may partly explain the observed increase in $T_N$ found here. The interplane coupling $J_c$ is of the same order as that previously noted in manganites confirming the largely two-dimensional behavior of h-LFO [25]. The $D_z$ value determined here is much smaller than for LuMnO$_3$ [20, 21, 23] but is close to the value obtained from *ab-initio* calculations for h-LFO [6]. Note that the selection between the $\Gamma_1$ and $\Gamma_2$ representations depends sensitively on the details of the local anisotropy tensor, so that this small $D_z$ may allow enhanced fluctuations between the two representations as observed above $T_N$. However, the magnetic ground state is well described by a single magnetic representation thatallows a net FM moment along the c-axis, corroborated by the onset of a small ferromagnetic moment below $T_N$. Above $T_N$ demonstrate both $\Gamma_1$ and $\Gamma_2$ representations are evident in the spin correlations, indicating that they are quite close in energy. This system therefore appears to be an ideal candidate to realize the novel magnetoelectric coupling recently predicted theoretically. Measurements of single crystals of this and similar bulk h-LFO systems then would be of tremendous interest in elucidating futher details about this system.

## Acknowledgements

Work at Rutgers University was supported by the DOE under Grant No. DE-FG02-07ER46382, , and work at Postech was supported by the Max Planck POSTECH/KOREA Research Initiative Program [Grant No. 2011-0031558] through NRF of Korea funded by MEST. This work utilized facilities supported in part by the National Science Foundation under Agreement No. DMR-0944772. S. M. Disseler acknowledges the support of the Nation Research Council NIST postdoctoral research associateship.

## References

[1] Sang-Wook Cheong and Maxim Mostovoy, Nature Materials **6** 14 (2007).

[2] J. Wang, J. B. Neaton, H. Zheng, V. Nagarajan, S. B. Ogale, B. Liu, D. Viehland, V. Vaithyanathan, D. G. Schlom, U. V. Waghmare, N. A. Spaldin, K. M. Rabe, M. Wuttig, R. Ramesh. Science **299** 1719 (2003).

[3] T. Katsufuji, M. Masaki, A. Machida, M. Moritomo, K. Kato, E. Nishibori, M. Takata, M. Sakata, K. Ohoyama, K. Kitazawa, and H. Takagi. Phys. Rev. B **66** 13434 (2002).

[4] E. Magome, C. Moriyoshi, Y. Kuroiwa, A. Masuno, and H. Inoue, Jpn. J. Appl. Phys. 49, 09ME06 (2010).

[5] Wenbin Wang, Jun Zhao, Wenbo Wang, Zheng Gai, Nina Balke, Miaofang Chi, Ho Nyung Lee, Wei Tian, Leyi Zhu, Xuemei Cheng, David J. Keavney, Jieyu Yi, Thomas Z. Ward, Paul C. Snijders, Hans M. Christen, Weida Wu, Jian Shen, and Xiaoshan Xu, Phys. Rev. Lett. **110** 237601 (2013).

[6] Hena Das, Aleksander L. Wysocki, Yanan Geng, Weida Wu, and Craig J. Fennie, Nature Communications **5** 2998 (2014).

[7] R. O. Cherifi, V. Ivanovskaya, L. C. Phillips, A. Zobelli, I. C. Infante, E. Jacquet, V. Garcia, S. Fusil1, P. R. Briddon, N. Guiblin, A. Mougin, A. A. Ünal, F. Kronast, S. Valencia, B. Dkhil, A. Barthélémy and M. Bibes. Nat. Mat. 13, 345 (2014).

[8] Jarrett A. Moyer, Rajiv Misra, Julia A. Mundy, Charles M. Brooks, John T. Heron, David A. Muller, Darrell G. Schlom, and Peter Schiffer. APL Materials **2** 012106 (2014).

[9] N.P. Cheremisinoff, Handbook of Ceramics and Composites (M. Dekker, New York, 1990).

[10]. Atsunobu Masunoa, Soichiro Sakaib, Yasutomo Araic, Hiroshi Tomiokad, Fumiaki Otsubod, Hiroyuki Inouea, Chikako Moriyoshib, Yoshihiro Kuroiwab and Jianding Yuc. Ferroelectrics **378** 169 (2009).

[11] Atsunobu Masuno, Atsushi Ishimoto, Chikako Moriyoshi, Naoaki Hayashi, Hitoshi Kawaji, Yoshihiro Kuroiwa, and Hiroyuki Inoue. Inorg. Chem. **52** 11889 (2013)

[12] J. W. Lynn, Y. Chen, S. Chang, Y. Zhao, S. Chi, W. Ratcliff, II, B. G. Ueland, and R. W. Erwin, Journal of Research of NIST 117, 61-79 (2012)

[13] J. Rodriguez-Cavajal, Physica B 192 55 (1993).

[14] N. Fujimura, T. Ishida, Y. Takeshi, and T. Ito. Appl. Phys. Lett. **69** 1011 (1996).

[15] S. Quezel, J. Rossat-Mignod, E. F. Bertaut, Solid State Comm **14** 941 (1974).

[16] P. J. Brown and T. Chatterji, J. Phys. Condens Matter **18** 10085 (2006).

[17] Joel S. Helton, Deepak K. Singh, Harikrishnan S. Nair, and Suja Elizabeth, Phys. Rev. B **84** 064434 (2011).

[18] A. Munoz, J. A. Alonso, M. J. Martinez-Lopez, M. T. Casais, J. T. Martinez, M. T. Fernandez-Diaz. Phys. Rev. B **62** 9498 (2000).

[19] B. E. Warren. Phys. Rev. **59** 693 (1941).

[20] W. Oh et. al, Phys. Rev. Lett. **111** 257202 (2013).

[21] T. J. Sato, S. -H. Lee, T. Katsufuji, M. Masaki, S. Park, J. R. D. Copley, and H. Takagi. Phys. Rev. B **68** 014432 (2003).

[22] S.-H. Lee, C. Broholm, W. Ratcliff, G. Gasparovic, Q. Huang, T. H. Kim and S.-W. Cheong. Nature **418** 856 (2002).

[23] H. J. Lewtas, A. T. Boothroyd, M. Rotter, D. Prabhakaran, H. Muller, M. D. Lee, B. Roessli, J. Gavilano, P. Bourges. Phys. Rev. B **82** 184420 (2010).

[24] S. Toth, B. Lake, arXiv:1402.6069

[25] O. P. Vajk, M. Kenzelmann, J. W. Lynn, S. B. Kim and S.-W. Cheong, Phys. Rev. Lett. **94**, 087601 (2005).

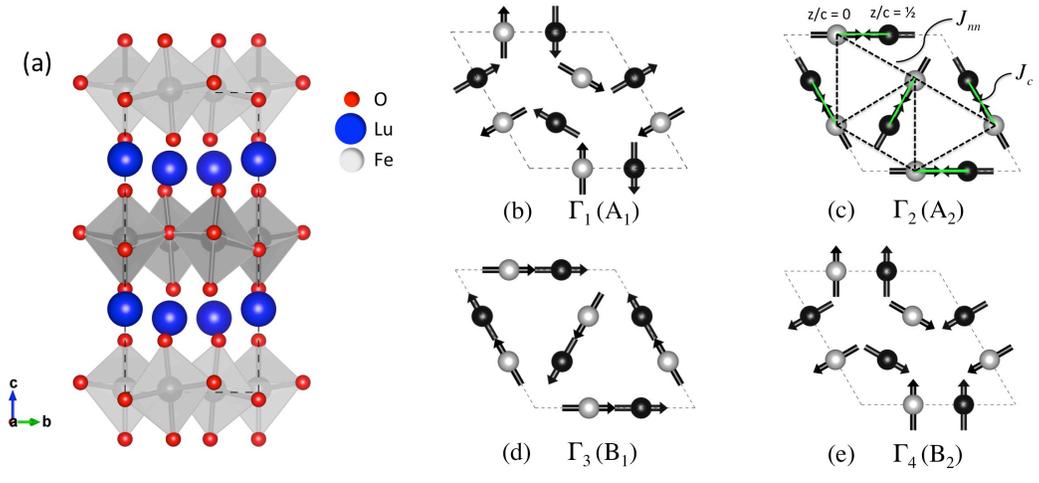

Figure 1: (Color online) (a) $P6_3cm$ crystal structure of h-LFO determined from NPD refinement. (b-e) Potential magnetic structures of h-LFO. The $\Gamma_i$'s correspond to the one-dimensional irreducible representations while labels in parenthesis correspond to the notation used in Ref. [6]. The exchange interactions $J_{nn}$ and $J_c$ are shown in (c).

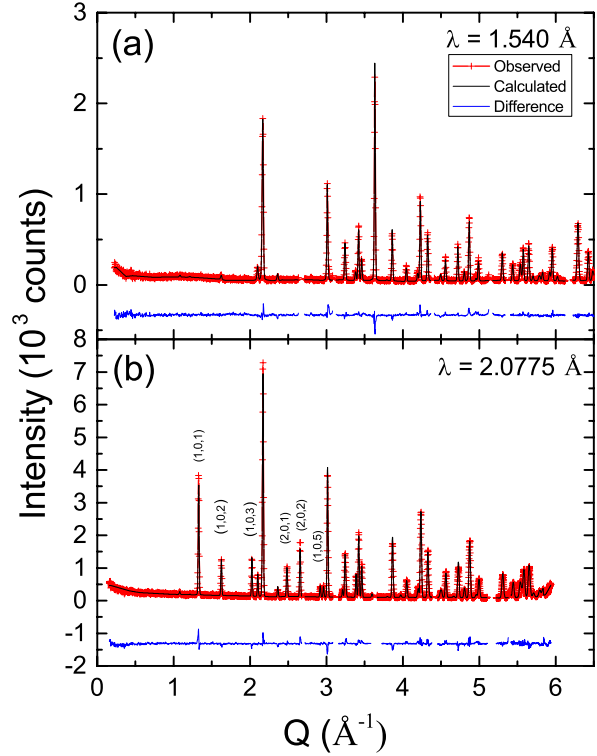

Figure 2: (Color online) Neutron powder diffraction patterns with Rietveld refinements at (a) 200 K ($\lambda$ = 1.540 Å) and (b) 5 K ($\lambda$ = 2.075 Å), where magnetic reflections appearing at low temperatures are labeled accordingly. The difference between the refined and observed spectra is shown offset in each panel respectively.

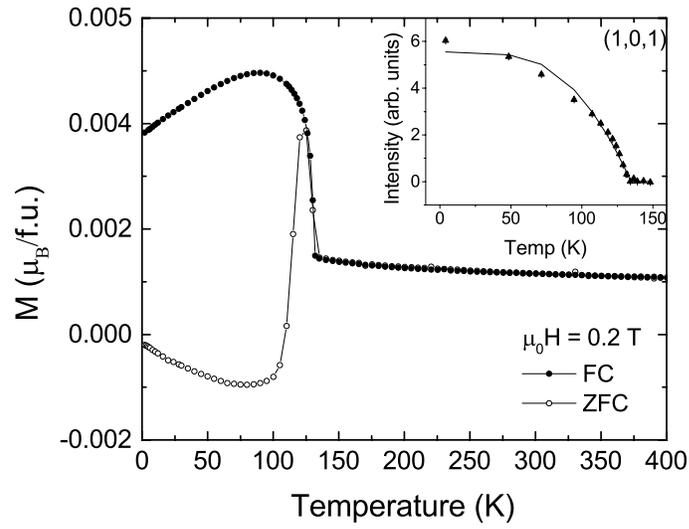

Figure 3: Temperature dependent magnetization taken on warming in a constant field after either FC (closed-circles) or ZFC (open-circles) the sample to 2 K. Inset: Integrated intensity of the magnetic (1, 0, 1) peak as a function of temperature. The solid curve is a mean-field fit to the intensity with $T_N = 134$ K.

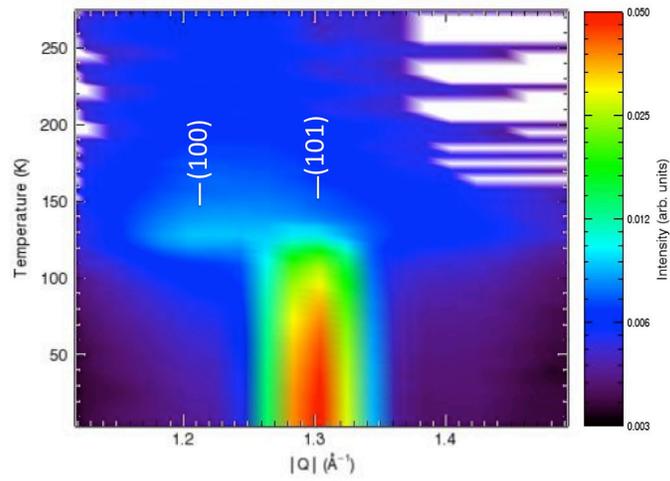

Figure 4: (Color online) Temperature dependence of the total scattering intensity measured on BT-7 using the position sensitive detector. The wavevectors corresponding to the (100) and (101) Bragg reflections are indicated for comparison.

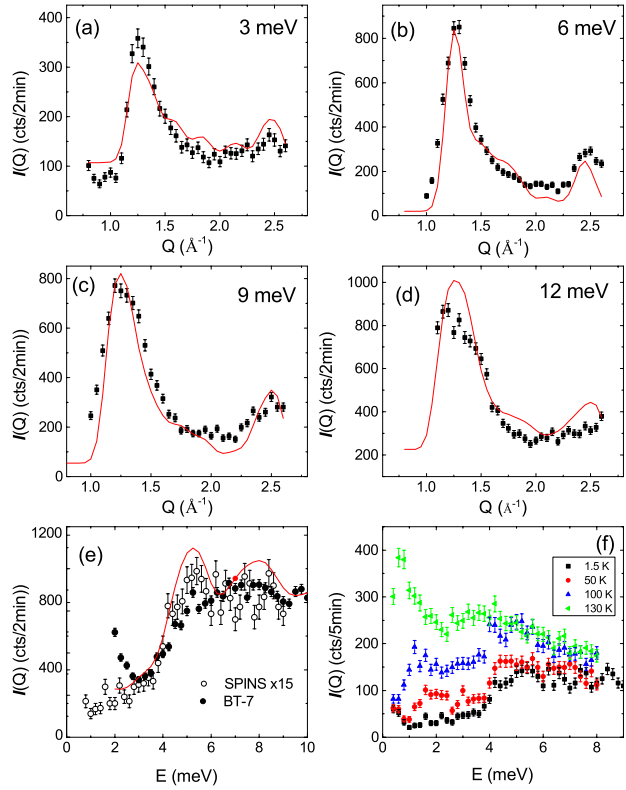

Figure 5: (Color online) Constant energy transfer scans at 3 K for $E =$ (a) 3 meV, (b) 6 meV, (c) 9 meV and (d) 12 meV. The solid (black) circles are the data and the solid (red) curve is the model fit as described in the text. (e) Constant $Q = 1.3$ Å data from SPINS and BT-7 where the SPINS data have been multiplied by a factor of 15 for comparison with BT-7. The solid curve is the fit to the BT-7 data as in (a)-(d). (f) Temperature dependent intensity for constant $Q = 1.3$ Å measured on SPINS.